\documentstyle[12pt]{article}    

\begin{document}

\title{
Probabilistic Quantum Teleportation}
\author{Pankaj Agrawal and Arun K. Pati$^{(1)}$\thanks{Emails: 
agrawal@iopb.res.in, akpati@iopb.res.in}\\
Institute of Physics, Bhubaneswar-751005, Orissa, India.\\
$^{(1)}$ Informatics, University of Wales, Bangor, LL57 1UT, UK.}

\newcommand{\Nol}{{1 \over \sqrt{1 + |\ell|^ 2} }}
\newcommand{\Non}{{1 \over \sqrt{1 + |n|^2    } }}
\newcommand{\Nop}{{1 \over \sqrt{1 + |p|^2    } }}
\newcommand{\Nolp}{{1 \over \sqrt{1 + |\ell^{\prime}|^ 2} }}
\newcommand{\Nom}{{1 \over \sqrt{1 + |m|^2    } }}
\newcommand{\Nopp}{{1 \over \sqrt{1 + |p^{\prime}|^2    } }}

\newcommand{\be}{\begin{equation}}
\newcommand{\ee}{\end{equation}}
\newcommand{\bea}{\begin{eqnarray}}
\newcommand{\eea}{\end{eqnarray}}

\newcommand{\rag}{\rangle}
\def\ie{\hbox{\it i.e.}{}}
\def\eg{\hbox{\it e.g.}{}}

\date{\today}
\maketitle
\def\ra{\rangle}
\def\la{\langle}
\def\ver{\arrowvert}

\begin{abstract}
We consider a generalized quantum teleportation protocol for an unknown qubit
using non-maximally entangled state as a shared resource. Without recourse
to local filtering or entanglement concentration, using standard Bell-state 
measurement and 
classical communication one cannot teleport the state with unit fidelity and
unit probability. We show that using non-maximally entangled measurements 
one can teleport an unknown state with unit fidelity albeit with reduced 
probability.
We also give a generalized protocol for entanglement swapping using non-maximally
entangled states.
\end{abstract}

\vfill



\makebox[\textwidth][c]{\bf I. Introduction }
\vspace{.1in}

Many of the quantum information processing protocols typically involve 
sending quantum states from sender to receiver using quantum and 
classical channels. Transmission of an intact unknown state from one place to 
another is very important in the field of quantum information.
One amazing discovery in this context is teleportation of an unknown
quantum state with the help of a maximally entangled channel, local
operation and classical
communications \cite{bbcjpw}. In standard teleportation protocol, Alice
performs a Bell-state
measurement on the unknown state and one-half of the maximally entangled 
pair and depending on the measurement outcome Bob applies a local unitary
operation to recover the unknown state. This has also been experimentally 
verified in recent years \cite{db,dbe,af}. The study of quantum teleportation 
protocol is not only limited to qubits and qudits (systems in $d$-dimensional
Hilbert spaces) but also to quantum systems in infinite dimensional Hilbert
spaces \cite{lv,bk}. Quantum teleportation can also be understood as a quantum
computation \cite{bb} and it has been even suggested that quantum teleportation
will play an important role as a primitive subroutine in quantum computation
\cite{dg}.

In real situations sender and receiver may not have shared maximally entangled
state but some form of non-maximally entangled pure state (due to some
imperfection at the source). Usually if one follows the standard
protocol, one will not be able to complete the teleportation process
with unit fidelity and unit probability. Rather, the fidelity will 
depend on the parameters of the the unknown state and the teleportation 
will not be reliable. Of course,
if one has several non-maximally entangled 
pairs one can first perform entanglement concentration \cite{bbps} and 
then recover fewer perfect maximally entangled pairs, and then use one of 
them to teleport an unknown state using the standard protocol.
If Alice and Bob have only  one pair, they can perform local 
filtering \cite{gis} first,
and convert a non-maximally entangled pair to maximally
entangled pair with certain probability. Then they can follow standard
protocol.

In this letter, we consider the question of teleporting an unknown state
with {\em unit fidelity} but less than unit probability when two parties share 
a non-maximally entangled state. We should
mention that there has been a proposal to teleport an unknown state using
any pure entangled state but using generalized measurements such as POVMs
\cite{tal}. This has been termed as conclusive teleportation.
Also, there has been a qubit assisted conclusive teleportation process 
\cite{band}. However, those protocols are different than ours as we will
see below. We provide a simple   
protocol that uses single shot standard orthogonal projections in {\em 
non-maximally entangled basis} and able to teleport an arbitrary state with
unit fidelity albeit less than unit probability, hence probabilistic
teleportation.
We discuss various special cases from probabilistic to deterministic 
teleportation of unknown states. Further, we generalize entanglement swapping
protocol for non-maximally entangled states.

\vspace{0.05in}
\makebox[\textwidth][c]{\bf II. Teleportation with non-maximally entangled
state}\\

In this section we present our simple scheme to teleport an unknown state
using non-maximally entangled state.
Let us consider two observers A and B (conventionally called Alice and Bob)
who share a pure non-maximally entangled state as a resource:

\be
|\Phi_{\rm{res}} \rag_{12} = \Non \,(|00\rag_{12} + \, n |11\rangle_{12}),
\ee
where $n$ is a known complex number. It is understood that
qubits `1' and `2' are with Alice and Bob, respectively. Notice that
because of the existence
of Schmidt decomposition \cite{ep,akp} any two qubit entangled state
$|\Phi \rag  \in  \cal{H}^{\rm 2} \bigotimes \cal{H}^{\rm 2}$ such as
\be
|\Phi \rag = a |00\rag +  b |11\rag + c|01 \rag + d |10\rag,
\ee
can be written as a superposition of two basis vectors. In general given
an arbitrary two-qubit state (2), the computational basis
$|00\rag$ and $|11\rag$ need not be the Schmidt basis, but
we assume that Alice and Bob know the Schmidt basis and coefficients. Then (1)
is the most general non-maximally entangled state up to local unitary
transformations relating Schmidt basis and computational basis states.
Now Alice receives a qubit in an unknown state
$|\psi \rag_{a} = (\alpha |0\rag_{a}  + \beta |1\rag_{a})$ with $ |\alpha|^2 +
|\beta|^2 = 1$. Alice wishes to teleport this state to Bob using the
non-maximally entangled resource, local operations and classical
communications (LOCCs).

In order to send the state, Alice will make a joint measurement on the
the two qubits: qubit `1' that is in the entangled state
$|\Phi_{\rm{res}}\rag_{12}$ with particle `2' and the other that
is in the state $|\psi\rag_{a}$. If Alice performs a measurement in the Bell
basis, then the state $|\psi\rag_{a}$ cannot be teleported faithfully, \ie,
with unit fidelity and unit probability.
However if the measurement is in a non-maximally entangled basis then it is
possible for Alice to send the state with
unit fidelity, though not with unit probability. 
Therefore we call our protocol probabilistic quantum teleportation.
We will also discuss how it is important to use
non-maximally entangled measurements having same amount of entanglement as
that of the shared resource. This is like taking out a nail by another nail!
To see this we carry out the following analysis.

First we give the most general set of basis vectors for two qubit
Hilbert space possessed by Alice. Since Alice can do whatever physical
operations within her laboratory we allow the general entangled basis states.
If we denote a set of basis vectors as $\{|00\rag, |01\rag,|10\rag,
|11\rag \}$, known as computational basis, then we can define another set of
mutually orthogonal basis vectors as:
\bea
        |\varphi^{+}_{\ell}\rag & = & \Nol \,(|00\rag + \,\ell \,|11\rag)  \\
        |\varphi^{-}_{\ell}\rag & = & \Nol \, (\ell^{*}|00\rag - \,|11\rag)
\eea

\bea
        |\psi^{+}_{p}\rag & = & \Nop \, (|01\rag + \,p \,|10\rag)  \\
        |\psi^{-}_{p}\rag & = & \Nop \,(p^{*}|01\rag - \,|10\rag)
\eea
Here $\ell$ and $p$ are complex numbers in general. 
We notice that when
$\ell = p = 0$, this basis reduces to the computational basis
which is not entangled.
For  $\ell = p = 1$, it reduces to the Bell basis which is maximally
entangled. Therefore this set interpolates between untangled and
maximally entangled set of basis vectors. Also note that the set
$|\varphi^{\pm}_{\ell}\rag$ and $|\psi^{\pm}_{p}\rag$ have different
amount of entanglement. As measured by von Neumann entropy \cite{pr},
the entanglement of
$E(|\varphi^{\pm}_{\ell}\rag)= (- \, L^2\rm{log}_{2}L^2 - L^2 \, |\ell|^2 \,
\rm{log}_{2}L^2 |\ell|^2)$ and of
$E(|\psi^{\pm}_{p} \rag)= (- \,P^2 \rm{log}_{2}P^2 - P^2 \, |p|^2 \,
\rm{log}_{2}P^2 |p|^2)$, respectively are different for these sets
with $L = \Nol$ and $P = \Nop$ are real numbers.
However, when $\ell = p$, then all basis vectors have identical von
Neumann entropy.

We can invert the above transformations and we see:

\bea
  |00\rag & = & \Nol \; ( |\varphi^{+}_{\ell}\rag + \,\ell |\varphi^{-}_{\ell}\rag   )       \\
  |11\rag & = & \Nol \; (\ell^{*}  |\varphi^{+}_{\ell}\rag - \,|\varphi^{-}_{\ell}\rag )    \\
  |01\rag & = & \Nop \; (|\psi^{+}_{p}\rag + \, p |\psi^{-}_{p}\rag  )        \\
  |10\rag & = & \Nop \; (p^{*}  |\psi^{+}_{p}\rag - \, |\psi^{-}_{p}\rag )
\eea

Using the non-maximally entangled basis states given in (3-6) we can rewrite the
combined state of the input and resource state as:

\bea
|\psi\rag_{a} |\Phi_{\rm{res}} \rag_{12} & = &  N \; (\alpha |0\rag_{a} + \, \beta |1\rag_{a})\,
(|00\rag_{12} +\, n |11\rag_{12})  \nonumber  \\
                & = & N \; (\alpha |00\rag_{a1} |0\rag_{2} +
\, \alpha \,n |01\rag_{a1}|1\rag_{2} + \,\beta |10\rag_{a1} |0\rag_{2}
+\, \beta \,n |11\rag_{a1} |1\rag_{2} ) \nonumber \\
                & = & N [\,|\varphi^{+}_{\ell}\rag_{a1} \, (L \, \alpha |0\rag_{2}
+ \,L\, n \, \beta \, \ell^{*}  |1\rag_{2} ) \nonumber \\
                &    &  \;\;\;\;\;\;\;\;\;\; + \;|\varphi^{-}_{\ell} \rag_{a1}
( L \,\ell \,\alpha |0\rag_{2} - \, n \, L \, \beta |1\rag_{2} ) \nonumber  \\
		&    & \;\;\;\;\;\;\;\;\;\; + \; |\psi^{+}_{p}\rag_{a1}
( P \,\beta \, p^{*} |0\rag_{2} + \, P \,\alpha \, n |1\rag_{2} ) \nonumber\\
                 & &  \;\;\;\;\;\;\;\;\;\; +  \; |\psi^{-}_{p}\rag_{a1} (-P \,\beta|0\rag_{2} +
\, P \,\alpha \,n \, p |1\rag_{2} ) ].
\eea
Here $N = \Non, L = \Nol$ and $P = \Nop$ are real numbers.  Above expression
is the most general way of rewriting an unknown state and two qubit entangled
state. We now wish to have faithful transportation with nonzero
probability. Let us consider several scenarios involving various
choices of the parameters $\ell$ and $p$, given the value of $n$. Choice is at the
disposal of Alice.

(i){\it Standard teleportation protocol:}
Let us choose $ \ell = {1 \over \ell^{*}} = p^{*} = {1 \over p} = n$. 
In this case $\ell, n$ and $p$ can be pure phases, \ie, complex numbers
of unit modulus. Then faithful teleportation is possible with 
unit fidelity and unit probability. This
is classic teleportation \cite{bbcjpw}. Bob can regenerate the state
$|\psi \rag_{a}$ by applying the local unitary transformation 
$\sigma_{0}=I$, $\sigma_{z}$, $\sigma_{x}$, or  $i\,\sigma_{y}$ on 
his qubit. These transformations will  correspond to
the Alice's result of measurement $|\varphi^{+}_{\ell}\rag$, 
$|\varphi^{-}_{\ell}\rag$,
$|\psi^{+}_{p}\rag$, or $|\psi^{-}_{p}\rag$ respectively. Alice can communicate
the results of her measurement to Bob using 2 cbits of information over a
classical channel. Then Bob with his knowledge of the shared resource state can
find out the unitary operation needed to convert the state of his qubit to 
$|\psi\rag_{a}$.
The unitary operations $\sigma_{x}$, $i\,\sigma_{y}$, and $\sigma_{z}$ correspond
to the rotation by $180^{0}$ around the $x, y$, and $z$ axis respectively. We
would also like to emphasize that in a slightly modified version of the
classic teleportation protocol, Bob does not have to know the shared
resource state; only Alice has to know the shared resource state. In this
modified version of the protocol, Alice will encode in two cbits the
unitary transformation (instead of the state she has got after the
measurement), that Bob has to apply on his qubit to complete the teleportation.

(ii){\it Probabilistic teleportation protocol:}
 If we make the choice $ \ell = n = p^{*}$, or $\ell = n =
{1 \over p}$, or $\ell^{*} = {1 \over n} = p$, or $ \ell^{*} =
{1 \over n} = {1 \over p^{*}}$, then for any of these choices,
reliable teleportation is possible for only two out
of four possible results of the measurement. An interesting observation here
is that the above choice of parameters refers to the situation where the basis
used for joint measurements and the resource state have the same amount
of quantum entanglement, namely,
$E(|\Phi_{\rm{res}} \rag)= (- \, N^2\rm{log}_{2}N^2 - N^2 \, |n|^2 \,
\rm{log}_{2}N^2 |n|^2)$.
For example, in the case of first
choice, when the outcome is $|\varphi^{-}_{\ell = n}\rag$, then the state at
Bob's hand will be $(\alpha |0\rag - \beta|1\rag)$ and when the outcome is
$|\psi^{+}_{p=n}\rag$, then the state at Bob's hand is
$(\beta |0\rag + \alpha|1\rag)$. Therefore, when Alice sends two classical
bits to Bob he will apply $\sigma_z$ in the former and $\sigma_x$ in the
later case to recover the unknown state with unit fidelity.
The total probability of this successful teleportation will be given by
\be
P_{\rm{succ}}= { 2 |n|^2 \over (1 + |n|^2)^2}.
\ee
Thus, we can say that using
$E(|\Phi_{\rm{res}} \rag)= (- \, N^2\rm{log}_{2}N^2 - N^2 \, |n|^2 \,
\rm{log}_{2}N^2 |n|^2)$ amount of entanglement and two classical bits Alice
{\em can teleport an unknown state with unit fidelity and probability 
given in (12)}. This is one of the main result of our letter.
We see that this probability goes from zero for untangled
$|\Phi_{\rm{res}} \rag$ to one-half for the maximally entangled resource
state. (Other one-half will come from the other two possible outcomes
of the measurement when the shared resource and joint measurement are
maximally entangled sates.) In this sense we can regard our protocol as a
generalized quantum teleportation protocol (GQTP) that goes from
probabilistic one to deterministic one.

Furthermore, we can amplify the probability statistically by repetitions.
We can say that the reciprocal of the {\em average} success probability
must be the number of repetitions $R$ that are required in order to
successfully (all the time) teleport an unknown state with unit fidelity.
We see that one shall need on the average
at least $R= { (1 + |n|^2)^2 \over |n|^2}$ repetitions to get a
faithful teleportation with unit probability. Therefore, if Alice and Bob share
$R E(|\Phi_{\rm{res}} \rag)$ pairs of non-maximally entangled state they can
successfully teleport an arbitrary state using local operation and $2R$ bits
of classical communication. We also notice that as the degree of entanglement
increases, the number of required repetitions decreases and becomes one
for maximally entangled states as expected. It becomes infinite
for the untangled resource state. Thus when no prior shared entanglement exist
howsoever many times one tries, it will be impossible to teleport an unknown
state with unit fidelity.

Our approach is similar to the filtering approach,
however, there is  one important difference. In the
filtering approach one cannot proceed with the
protocol if the filtering does not succeed.
In principle, the unknown qubit can be put into memory until
the next entangled pair is sent, and so on until a successful filtering
event happens. Then one can proceed with the standard teleportation protocol.
Such an option is not available in our protocol as it is
not till the actual measurement of the qubit has taken place that success
or failure is known, and by that time the unknown qubit has been destroyed.
Therefore, we also need R copies of the unknown state in order to
get faithful (unit fidelity and unit probability) teleportation.
Since the state is provided by a third party (say Victor), who knows
the state, there is no `cost' involved as he can make one copy
or several of them.

There are also four different choices of parameters when only one out of four
results of the measurement would lead to faithful teleportation.
This choice of parameter values is given by
$\ell = {1 \over n^{*}}$, or $\ell = n$, or $p = n^{*}$, or $ p =
{1 \over n}$. In this case the total probability of
faithful teleportation will be ${ |n|^2 \over (1 + |n|^2)^2}$. This is
half of the scenario given above.
We note that unlike in the case of standard teleportation (where only
Alice has to know), in probabilistic teleportation both Alice and Bob 
have to know the shared
resource state. Only then, Bob will know what basis Alice has used for making
the measurement after he receives classical communication.

(iii){\it No teleportation:} If the values of $p$ and $\ell$ are not
related with that of
$n$, then teleportation is not possible with unit fidelity. This brings out
an interesting point: in order that an arbitrary  quantum entangled channel
is useful for teleportation {\em we must use an entangled measurement
containing the same amount of entanglement as the shared resource state.}
Even though shared entanglement is regarded as a resource and local 
entanglement is not (as Alice can create or destroy entanglement), 
still the above point is worth observing.

Before ending this section 
it may be noted that present experiments have reported teleportation 
of qubit with certain success probability less than unity, in spite of 
the sharing of maximally
entanglement \cite{db}. This limitation is a practical limitation on
Bell-state detection \cite{lcs}. Though our protocol behaves in a 
similar way to the
experiments in the sense that teleportation is only successful for a 
limited subset of the possible measurement results, fundamentally they 
are different.

\vspace{0.2in}

\makebox[\textwidth][c]{\bf III. Entanglement swapping with
non-maximally entangled states}

Another important prediction of quantum theory is that conditional upon
suitable joint measurement, two particles can be found in an entangled state
that have never interacted in the past. If there are two maximally
entangled pairs
then making a Bell measurement on two halves makes other two halves
maximally entangled. This is known as entanglement swapping \cite{marek,bose}.
In this section we will discuss how to generate entanglement between two
independent particles using non-maximally entangled states as the starting
resource and non-maximal measurements. Let us consider two pairs of qubits 
`$ab$' and `$12$'. Let them be in non-maximally entangled states 
$ |\varphi \rag_{ab}$ and  $|\psi \rag_{12}$. Here,

\bea
|\varphi \rag_{ab} &  = & M \,(|00\rag + \, m |11\rangle)_{ab},  \nonumber \\
|\psi \rag_{12} & = & N \, (|01\rag + \, n |10\rangle)_{12}
\eea

If an observer, Alice,  makes a measurement on the qubit pair `$a1$',
then we wish to analyze the state of the particle `$b$'
at Bob's location and the particle `$2$' at Charlie's location.
 For this, we consider the state of the combined system:

\bea
   |\varphi \rag_{ab} |\psi \rag_{12} & = & M \, N (|00\rag
+ \, m |11\rangle)_{ab} \, (|01\rag + \, n |10\rangle)_{12}  \nonumber \\
 & = & M \, N [ L\, P^{\prime} \, (|\varphi^{+}_{\ell}\rag_{a1} +
\, \ell |\varphi^{-}_{\ell}\rag_{a1}  )
       (|\psi^{+}_{p^{\prime}}\rag_{b2} + \, p^{\prime} |\psi^{-}_{p^{\prime}}\rag_{b2} ) \nonumber \\
          & & \;\;\; + \, m \, n \, L \, P^{\prime} \, (\ell^{*}|\varphi^{+}_{\ell} \rag_{a1} -
	                         \,|\varphi^{-}_{\ell}\rag_{a1} )
( p^{\prime *} |\psi^{+}_{p^\prime}\rag_{b2} - \,|\psi^{-}_{p^{\prime}}\rag_{b2} )   \nonumber \\
& & \;\;\;           +\, n \, P \, L^{\prime} \, (\, |\psi^{+}_{p} \rag_{a1} +
	                         \, p |\psi^{-}_{p}\rag_{a1} )
   (|\varphi^{+}_{\ell^\prime }\rag_{b2}  + \, \ell^{\prime} |\varphi^{-}_{\ell^{\prime}}\rag_{b2} )   \nonumber  \\
& &  \;\;\; + \, m\, P \, L^{\prime} \, (p^{*} \, |\psi^{+}_{p}\rag_{a1} - \, |\psi^{-}_{p}\rag_{a1}  )
       (\ell^{\prime *} \, |\varphi^{+}_{\ell^{\prime}}\rag_{b2} - \,|\varphi^{-}_{\ell^{\prime}}\rag_{b2} )] \nonumber \\
& = & M\, N [L \, P^{\prime}\, |\varphi^{+}_{\ell}\rag_{a1} \{|\psi^{+}_{p^{\prime}}\rag_{b2} ( 1 + m \,
                           n \, p^{\prime *} \ell^{*}) +|\psi^{-}_{p^\prime}\rag_{b2} ( p^{\prime} - m \,
                           n \, \ell^{ *} )
			     \}  \nonumber \\
 & & + L \, P^{\prime} \,|\varphi^{-}_{\ell}\rag_{a1} \{|\psi^{+}_{p^{\prime}}\rag_{b2} ( \ell - \,m \,
                           n \, p^{\prime *}) + \, |\psi^{-}_{p^{\prime}}\rag_{b2} ( p^{\prime} \ell + m \,
                           n  )  \}  \nonumber \\
& & +  P \, L^{\prime} \,|\psi^{+}_{p}\rag_{a1} \{|\varphi^{+}_{\ell^{\prime}}\rag_{b2} ( n + \, m\, p^{*} \,
                           \ell^{\prime *}) + |\varphi^{-}_{\ell^{\prime}}\rag_{b2} ( n \, \ell^{\prime} - m \,p^{*} ) \}  \nonumber \\
& & +  P \, L^{\prime} \,|\psi^{-}_{p}\rag_{a1} \{|\varphi^{+}_{\ell^{\prime}}\rag_{b2} ( n \, p - \,m \,
                           \ell^{\prime *}) + |\varphi^{-}_{\ell^{\prime}}\rag_{b2} ( n \, p \, \ell^{\prime } + \, m)  \} ] \nonumber \\
\eea

   Here $N = \Non,\, M = \Nom, \, L = \Nol,\, P = \Nop,\, L^{\prime} = \Nolp$, and $P^{\prime} = \Nopp$
   are real numbers. We are given the parameters $m$ and $n$, and we can choose $\ell, \ell^{\prime},
   p$ and $p^\prime$. In rewriting the four particle state above, we use the basis
   $\{ |\varphi^{\pm}_{\ell}\rag, \,|\psi^{\pm}_{p}\rag \}$ for pair `$a1$',
   while for the pair `$b2$' the basis $\{ |\varphi^{\pm}_{\ell^\prime}\rag,
   \, |\psi^{\pm}_{p^\prime}\rag \}$ is used.

   For faithful entanglement swapping parameters $\ell, p, \ell^{\prime}$, and
   $p^{\prime}$ must satisfy a set of conditions. 
As an illustration, we choose the following set of conditions: (1)
   $p^{\prime} = \, m \, n \, \ell^{*}$ {\em and} $\ell = \, m \, n \, p^{\prime *}$
   for  $p^{\prime}$ and $\ell$; (2)
   $n \, \ell^{\prime} = m \, p^{ *}$ {\em and} $n \, p = \, m \, \ell^{\prime *}$ for
    $\ell^{\prime}$ and $p$.  A different set of conditions should lead to similar conclusions.  We now consider following situations:

(i){\it Standard entanglement swapping:}
First, we state the conditions under which standard entanglement swapping
is possible.
If we choose $ \ell = {1 \over \ell^{*}} = p^{*} = {1 \over p} =
\ell^{\prime} = {1 \over \ell^{\prime *}}
= p^{\prime } = {1 \over p^{\prime *}} = m = n$, then in this case faithful
swapping is possible with unit probability. And all parameters are pure
phases and all the considered entangled
states are maximally entangled. However, we note that only $m$ and $n$ need
be pure phases, \ie, two initial pairs must be maximally entangled.
Measurement basis need not be Bell basis. It can be non-maximally
entangled basis with the requirement:
$|\ell| = |p^{\prime}|$ and $|p| = |\ell^{\prime}|$. The resulting state at
Bob and Charlie's location will be non-maximally entangled. Note that if the observer
for the pair `$a1$' and the pair `$b2$' use the same basis, then this means
$\ell = p$, and all basis vectors have same degree of entanglement.

   (ii) {\it Probabilistic entanglement swapping:}
 Two conditions (1) and (2) given above cannot be satisfied simultaneously
  if the two initial
       states are not maximally entangled, \ie, when $m$ or $n$ are not pure
       phases. In such a case two out of four measurements will lead to
       reliable entanglement swapping. There are many possible
       choices for the values of the parameters that would lead to the
       reliable swapping but with probability less than unity. One such choice
       is $ \ell = \, {1 \over n^{*}}, \, p^{\prime} = \, m, \, p = \, {1 \over m^{*}}$,
        and $\ell^{\prime} = \, {1 \over n } $. In this
       case successful swapping probability will be given by
\be
P_{\rm {succ}}= M^{4} \, N^{4} \,
       [ |n|^2 \, (1 + |m|^{2})^2  + \, |m|^2 ( 1 + |n|^{2} )^2  ].
\ee
This reduces to one-half, when the two initial states are maximally entangled
 (and other half will come from the two other outcomes).

 There is an interesting possibility if the entanglement of the two initial states
 is not maximal, but identical. This happens when $|m| = {1 \over |n|}$ or
 $|m| = |n|$. In this case three out of four possible measurement results would
 lead to reliable entanglement swapping and the swapping probability will be:

\be
P_{\rm {succ}}^{\prime} = 3 \, |n|^{2} \, N^{8} \, (1 + \, |n|^2)^2
\ee

 For lesser constraints on the parameter values, as in the case of a
qubit state teleportation, only one out of four possible measurement
results will lead to faithful entanglement swapping. An example of these
parameter values is $ \ell = \, {1 \over n^{*}}, \, p^{\prime} = \, m $.
And there are no constraints on other parameter values.

(iii) {\it No swapping:} In this most general case when the parameters
values are not related to original resource, entanglement swapping is not
possible.

Thus the scheme presented here tries to capture probabilistic and deterministic
entanglement swapping protocols for qubits.
Since entanglement swapping can be understood as a teleportation of
an entangled state, we have generalized to such scenarios as well.

\makebox[\textwidth][c]{\bf IV. Conclusions}

In conclusion,
we have shown that it is possible to teleport an unknown state with unit
fidelity but less than unit probability using non-maximally entangled states.
The difference between the present protocol and the existing ones is that
neither we use local filtering nor entanglement concentration and then follow the
standard teleportation protocol. It is a {\em single shot teleportation
protocol} for non-maximally entangled resource without first converting
to a maximally entangled pair.
The key to this generalization is if one uses non-maximally
entangled state as a resource use non-maximally entangled-state measurement
containing same amount of entanglement as that of the shared resource
instead of the Bell measurement. This also points, perhaps, to a link between
global and local entanglement. In some sense ours is a generalized
quantum teleportation protocol that encompasses in a simple way
probabilistic as well as deterministic (standard) teleportation protocols.
In addition we have presented a scheme how to perform entanglement 
swapping using these resources.
In future it will be interesting to extend these probabilistic teleportation
schemes for higher dimensional Hilbert space and continuous variable
systems. We hope that with the existing technology it may be possible to
implement the probabilistic quantum teleportation protocol with ease. 


\end{document}